\documentclass{article}

\usepackage{PRIMEarxiv}

\usepackage[utf8]{inputenc} 
\usepackage[T1]{fontenc}    
\usepackage{hyperref}       
\usepackage{url}            
\usepackage{booktabs}       
\usepackage{amsmath,amssymb,amsfonts}
\usepackage{nicefrac}       
\usepackage{microtype}      
\usepackage{lipsum}
\usepackage{fancyhdr}       
\usepackage{graphicx}       
\graphicspath{{media/}}     

\usepackage{cite}

\usepackage{algorithmic}

\usepackage{textcomp}

\let\labelindent\relax
\usepackage{enumitem}
\usepackage{pifont}
\usepackage{siunitx}
\usepackage{multirow}

\usepackage{mathrsfs}

\makeatletter

\newcommand{\subtext}[2]{#1_{\text{#2}}}
\newcommand{\perc}[1]{\SI{#1}{\percent}}

\title{Unsupervised reconstruction of accelerated cardiac cine MRI using Neural Fields
\thanks{\textit{\underline{Citation}}: 
\textbf{Authors. Title. Pages.... DOI:000000/11111.}} 
}

\begin{document}


\begin{flushleft}

\hsize\textwidth
\linewidth\hsize
\vskip 0.1in

\hrule height 2\p@
\vskip 0.25in
\vskip -\parskip%

\centering
{\LARGE
\textbf\newline{Unsupervised reconstruction of accelerated cardiac cine MRI using Neural Fields\par}
}

\vskip 0.29in
\vskip -\parskip
\hrule height 2\p@
\vskip 0.09in%

\vskip 0.1in 

Tabita Catal\'an\textsuperscript{1},
Mat\'ias Courdurier\textsuperscript{2},
Axel Osses\textsuperscript{3,4},
Ren\'e Botnar\textsuperscript{5,7},
Francisco Sahli Costabal\textsuperscript{5,6,7*,$\sharp$},
Claudia Prieto\textsuperscript{2,6,7,$\sharp$}
\\
\bigskip
\bf{1} Millennium Nucleus for Applied Control and Inverse Problems, Santiago, Chile.
\\
\bf{2} Department of Mathematics, Pontificia Universidad Católica de Chile, Santiago, Chile.
\\
\bf{3} Center for Mathematical Modeling, Santiago, Chile 
\\ 
\bf{4} Department of Mathematical Engineering, Universidad de Chile, Santiago, Chile.
\\
\bf{5} Institute for Biological and Medical Engineering, Pontificia Universidad Católica de Chile, Santiago, Chile.
\\ 
\bf{6} School of Engineering, Pontificia Universidad Católica de Chile, Santiago, Chile 
\\
\bf{7} Millennium Institute for Intelligent Healthcare Engineering, Santiago, Chile
\\ 

\bigskip

*: fsc@ing.puc.cl, $\sharp$: equal contribution.

\end{flushleft}

\begin{abstract}

Cardiac cine MRI is the gold standard for cardiac functional assessment, but the inherently slow acquisition process creates the necessity of reconstruction approaches for accelerated undersampled acquisitions. Se\-veral regularization approaches that exploit spatial-temporal redundancy have been proposed to reconstruct undersampled cardiac cine MRI. More recently, methods based on supervised deep learning have been also proposed to further accelerate acquisition and reconstruction. However, these techniques rely on usually large dataset for training, which are not always available. In this work, we propose an unsupervised approach based on implicit neural field representations for cardiac cine MRI (so called NF-cMRI). The proposed method was evaluated in \textit{in-vivo} undersampled golden-angle radial multi-coil acquisitions for undersampling factors of 26x and 52x, achieving good image quality, and comparable spatial and improved temporal depiction than a state-of-the-art reconstruction technique.
\end{abstract}


\section{Introduction}
\label{sec:introduction}


Magnetic Resonance Imaging (MRI) is an essential tool for cardiac functional assessment \cite{mojibianChapterCardiacMagnetic2018}. Dynamic cardiac cine MRI is employed to obtain comprehensive information throughout the cardiac cycle. Cardiac cine imaging is conventionally performed synchronizing the acquisition with the patient's electrocardiogram and under breath-hold, in order to minimize cardiac and respiratory motion within each frame. Nevertheless, this approach limits the scan time, leading to spatial and temporal resolution constraints.


Several approaches have been proposed to accelerate cardiac MRI acquisition by obtaining less samples than those established by Nyquist (i.e. undersampling) \cite{donevaMathematicalModelsMagnetic2020}, \cite{knollDeepLearningMethodsParallel2020}. Undersampling introduces aliasing in the image, thus to preserve image quality it is necessary to use regularized reconstruction techniques, exploiting prior information and/or spatial-temporal redundancies. Parallel Imaging (PI)\cite{pruessmannSENSESensitivityEncoding1999}, \cite{griswoldGeneralizedAutocalibratingPartially2002} and Compressed Sensing (CS) \cite{lustigCompressedSensingMRI2008} are the most common and extensively utilized methods in this regard. PI exploits the spatial sensitivity encoding provided by multiple receiver coil acquisitions. CS capitalizes on the sparsity exhibited by images under specific transformations, such as the temporal Fourier transform for dynamic images. CS and PI serve as the fundamental building blocks for a wide array of approaches in cardiac cine reconstruction \cite{menchon-laraReconstructionTechniquesCardiac2019}. An example of such approach is GRASP \cite{fengGoldenangleRadialSparse2014}, which combines PI and CS for golden-angle radial sampling acquisitions.

Methods based on machine and deep learning have also been proposed to further accelerate MRI acquisition and reconstruction. These approaches are usually divided according to the task performed in: image-to-image denoising \cite{koflerSpatioTemporalDeepLearningBased2020a}; direct mapping from acquired k-space to the reconstructed image \cite{zhuImageReconstructionDomaintransform2018}; physics-based k-space learning \cite{akcakayaScanspecificRobustArtificialneuralnetworks2019} and unrolled optimizations \cite{schlemperDeepCascadeConvolutional2018}, and combination of those \cite{hammernikPhysicsDrivenDeepLearning2023}. The majority of these undersampled reconstruction methods follow a supervised paradigm, relying on extensive databases for model training. However, supervised approaches present some challenges. First, large k-space datasets for training are not always available.  Performance degradation has been reported when encountering variations in anatomical regions \cite{hammernikSystematicEvaluationIterative2021}, magnetic field intensities, scanner manufacturer, etc. \cite{muckleyResults2020FastMRI2021}, limiting the generalization of these techniques. Furthermore, these methods have been described as unstable, as subtle and imperceptible modifications in either the image or k-space domain can exert significant impacts on the resulting reconstructed image, such as creating hallucinations or image-like artifacts that are difficult to discard as nonphysical \cite{antunInstabilitiesDeepLearning2020}.

Recently, unsupervised approaches have been proposed to overcome these challenges. Untrained Convolutional Neural Networks (CNN) have been used successfully as priors in various imaging problems such as superresolution, inpainting and denoising \cite{ulyanovDeepImagePrior2020}, \cite{qayyumUntrainedNeuralNetwork2023}. This technique is commonly known as Deep Image Prior (DIP), and has been applied in MRI reconstruction \cite{zhaoReferenceDrivenCompressedSensing2020}, \cite{aggarwalENSUREGeneralApproach2023}, and in dynamic MRI reconstruction in particular \cite{yooTimeDependentDeepImage2021}, \cite{zouDynamicImagingUsing2021a}.



Neural fields \cite{xieNeuralFieldsVisual2022} is an emerging research field that centers around coordinate-based neural networks, also known as implicit neural representations. These networks employ fully connected Multi Layer Perceptrons (MLPs) to approximate fields with low-dimensional input and output. In this approach, the input of the neural network is a coordinate, and the expected output corresponds to the field value at that coordinate. 

This type of network suffers from spectral bias \cite{rahamanSpectralBiasNeural2019}, \cite{basriConvergenceRateNeural2019}, i.e. low frequencies are learned much earlier than high frequencies resulting in blurry and detail-lacking outputs. Various techniques have been proposed to address this issue. These include using periodic activation functions like SIREN \cite{sitzmannImplicitNeuralRepresentations2020} or architectures such as BACON \cite{lindellBaconBandlimitedCoordinate2022} that increase recoverable frequencies with additional layers, as well as preprocessing input coordinates through higher-dimensional mapping, such as hash-encoding \cite{mullerInstantNeuralGraphics2022} or Gaussian Fourier Features \cite{tancikFourierFeaturesLet2020a}.

 Implicit neural representations methods have been recently proposed for medical image reconstruction. For instance, NeRP \cite{shenNeRPImplicitNeural2022} utilizes implicit neural representations to reconstruct static data from Computed Tomography (CT) and MRI. This approach requires a prior image from the same scanner and patient. NeSVOR \cite{xuNeSVoRImplicitNeural2023} employs neural fields for 3D brain reconstruction using 2D MRI slices. 

In this work, we propose a novel unsupervised approach for cardiac cine MRI undersampled reconstruction with golden-angle radial multiple-coil acquisition. This approach, so-called NF-cMRI, relies on implicit regularization provided by spatio-temporal Fourier Features in a implicit neural representation. Unlike conventional supervised methods, the proposed approach does not require a large training database, as it can be trained on a single acquisition. 

NF-cMRI relies on a coordinate-based network that we called ``Intensity Net''. This network takes spatio-temporal coordinates as input and pre-processes them using a modified Gaussian Fourier Features technique, treating temporal coordinates differently from spatial ones. These coordinates are then passed through an MLP to generate the reconstructed intensity at any spatial-temporal coordinate. In other words, Intensity Net provides a continuous representation of the reconstructed image, parametrized by the MLP weights. The inverse problem of image reconstruction is solved iteratively in the training of the Intensity Net, by optimizing the parameters of the MLP. The resulting image is subsequently weighted with the corresponding coil sensitivity maps, and then compared to the acquired k-space data after applying the Fourier transform to ensure data consistency. For the Fourier transform, we make use of the Fourier Slice Theorem, which allows for a simple continuous implementation based on Radon transform without the need of the non-uniform fast Fourier transform (NUFFT). A diagram of the whole process can be seen in Figure \ref{fig:training}.

The principal contributions of this work are summarized as follows: 
 (1) We propose NF-cMRI, an unsupervised reconstruction approach for cardiac cine MRI based on implicit neural representations, and test it on real data. Although there are previous works on MRI reconstruction using neural fields, they were mainly proposed for static images or evaluated on synthetic data. The proposed approach uses a modified Gaussian Fourier Features encoding, treating temporal coordinates differently from spatial ones, and is evaluated in \textit{in-vivo} complex-valued, multi-coil radial k-space data.
    (2) We use a simple spoke-based batch training strategy tailored for radial acquisitions, eliminating the need for NUFFT while capitalizing on the continuous nature of the proposed MLP representation. 

\section{Methods}\label{methods}


\begin{figure*}[!t]
\centerline{\includegraphics[width=\textwidth]{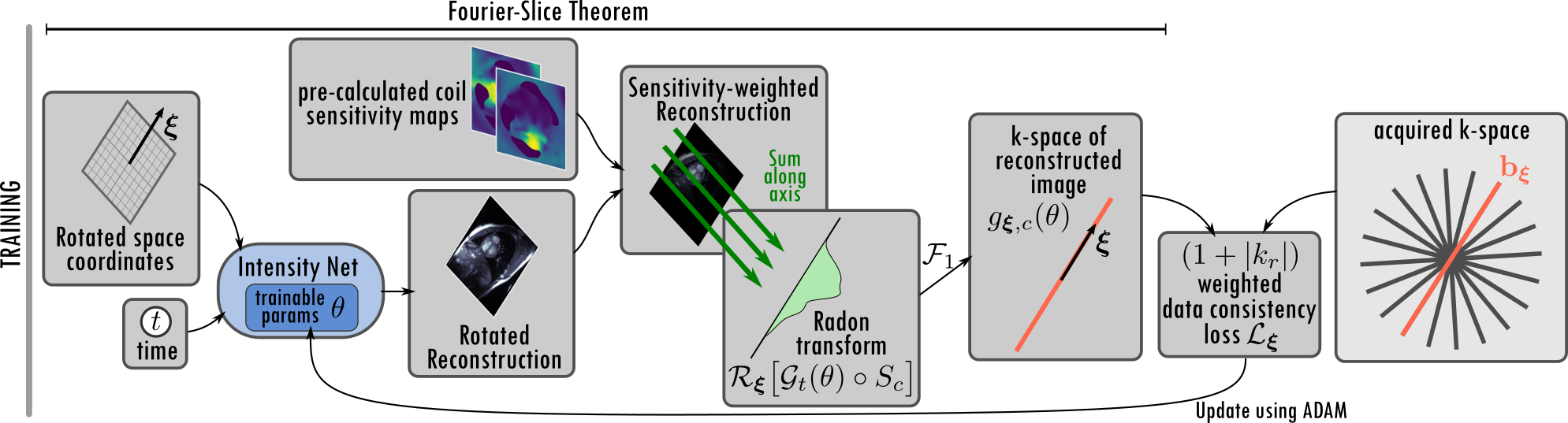}}
\caption{Overview of the self-supervised training strategy. A reconstruction is generated evaluating Intensity Net, a Network based on Neural Fields, on a set of coordinates. This reconstruction depends on trainable parameters \(\theta\). Coil sensitivity maps are pre-calculated. The k-space of the reconstruction is obtained using the Fourier-Slice Theorem, leveraging the continuous nature of Intensity Net to provide interpolation. This is compared with the acquired k-space with a weighted loss, and this value is used to update the trainable parameters in an iterative fashion.}
\label{fig:training}
\end{figure*}

\begin{figure}[!t]
    \centerline{\includegraphics[width=0.65\textwidth]{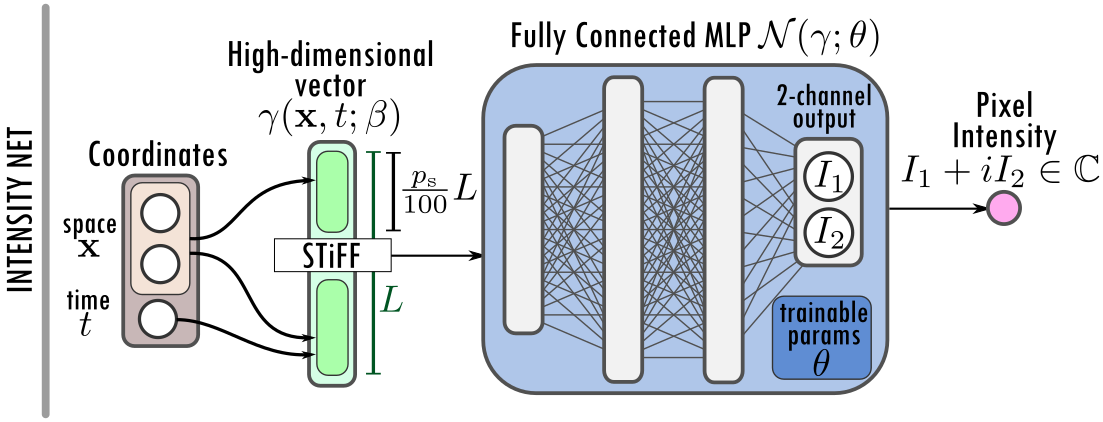}}
    \caption{Intensity Net is a coordinate-based MLP with a preprocessing step called Spatio-Temporal Fourier Features (STiFF). STiFF maps input coordinates to a high dimensional vector, improving the learning velocity of high frequencies. Space and time coordinates are treated differently by STiFF; a \(\subtext{p}{s}\perc{}\) of the entries of the high-dimensional vector only depends on space coordinates. The MLP has a 2-channel output that is combined into a single complex number.}
    \label{fig:intensity-test}
\end{figure}


Undersampled acquisition in MRI can be represented by applying a linear operator \(E\) to the image pixel values \(\mathbf{i}\). The encoding operator \(E\) encompasses coil-sensitivity weighting, Fourier transform and undersampling in the k-space. The ill-posed inverse problem of recovering the image pixel values \(\mathbf{i}\) from undersampled, noisy, measurements \(\mathbf{b} = E\mathbf{i} + \varepsilon\) is typically solved using a regularized reconstruction defined in equation \eqref{eq:regularized-prob}, with \(\mathcal{R}(\mathbf{i})\) being the regularization term and \(\lambda\) the hyperparameter controlling the strength of the regularization. 

\begin{equation}
     \min_{\mathbf{i}} \frac{1}{2} \| \mathbf{b} - E\mathbf{i}\|^2 + \lambda \mathcal{R}(\mathbf{i})
    \label{eq:regularized-prob}
\end{equation}



However, it is possible to use \textit{implicit} regularization  \cite{yooTimeDependentDeepImage2021,ulyanovDeepImagePrior2020}, where the image is represented by a function chosen from a set of prior functions with desirable qualities. When the set of candidate functions consists of images generated by neural networks $\mathcal{G}(\theta)$ with learnable parameters \(\theta\), the inverse problem can be recast as finding the best parameters rather than pixel values:

\begin{equation}
    \min_{\theta} \frac{1}{2} \| \mathbf{b} - E\mathcal{G}(\theta)\|^2 
    \label{eq:implicit-tddip-reg}
\end{equation}

In this work, we propose to define the set of prior functions as neural networks with a particular spatio-temporal structure that is suitable for cine MRI reconstruction. Each step of the proposed NF-cMRI is described below.


\noindent \textbf{Image reconstruction with neural fields.} The neural-field-based image reconstruction is done with Intensity Net, which is an MLP with a preprocessing step, as seen in Figure \ref{fig:intensity-test}. 
In the neural fields approach, an image of \(N_x\times N_y\) pixels is seen as the evaluation of an intensity function \(I\) on equispaced grid \(G = \{(x_i,y_j)\}_{i = 1, \dots, N_x, j = 1, \dots, N_y} \subseteq \mathbf{X}:=[-1,1]^2\). In this context, a neural field is  a coordinate-based MLP \(\mathcal{N}\) that approximates the function \(I\). The training process consists of finding weights \(\theta\) of the MLP such that \(\mathcal{N}(x,y;\theta) = I(x,y)\) for all \((x,y) \in G\). 

In a simple case where the image is known, the MLP can be trained using the values of the known image at the pixel positions as supervision points. After training, the MLP can represent a single image, and needs retraining for any new case. A well known issue of this approach is that MLPs have a spectral bias \cite{rahamanSpectralBiasNeural2019} when trained with gradient descent methods, namely, in the training process, the low frequency components of the data are learned much faster than high frequency ones, resulting in blurry images. Gaussian Fourier Features is an established technique to tackle this issue. This approach applies preprocessing to the coordinates, mapping the low-dimension input into a high-dimensional space. Specifically, a coordinate \(\mathbf{x} \in \mathbb{R}^d\) is mapped to the vector \(\gamma(\mathbf{x})\) by the Gaussian Fourier Feature mapping defined as 

\begin{equation}
\gamma(\mathbf{x}) =
\begin{bmatrix}
\cos(2\pi B\mathbf{x}) \\
\sin(2\pi B\mathbf{x}) \\
\end{bmatrix} \in \mathbb{R}^{2M}.
\label{eq:gff}
\end{equation}

The matrix \(B \in \mathbb{R}^{M \times d}\) encodes possible frequencies appearing in the image and it is fixed before training. Each component is sampled from a normal distribution, \(b_{ij} \sim N(0, \sigma)\). The hyper-parameters \(M\) and \(\sigma\) require tuning and depend on the size of the coordinate space and the frequency content of the image. The higher the value $\sigma$, the more high frequency components can be reconstructed, but excessively high values of $\sigma$ lead to the emergence of undesirable high-frequency artifacts. The mapping size \(M\) is significantly larger that \(d\). The vector \(\gamma(\mathbf{x})\) is used as input for the MLP. 

In order to extend neural fields to cardiac cine MRI undersampled reconstruction, some modifications must be made. MR images are complex, thus the MLP needs to include a 2-channel -- real and imaginary-- output. Furthermore, the reference fully sampled image is unavailable, as the aim is to reconstruct an image from undersampled k-space data. Additionally, applying Gaussian Fourier Feature to dynamic images is not straightforward. The frequencies observed in the spatial and temporal domains are inherently different, as the former represents the changes in tissue intensity and the latter represents the cardiac motion. This is also reflected in the resolution of the images, with e.g. 200-400 pixels per spatial dimension vs 15-30 time frames per cardiac cycle. This suggests that time cannot be treated as an additional spatial dimension, as the frequencies sampled in matrix $B$ would come from the same distribution for space and time.

To address this issue, here we propose treating temporal coordinates differently from spatial ones and assume a nearly-periodic cardiac cycle. We construct a new encoding, the Spatio-Temporal Fourier Features (STiFF), to achieve this goal:

\begin{equation}
\gamma(\mathbf{x}, t; \beta) =
\begin{bmatrix}
\cos(2\pi B_{\text{s}}\mathbf{x}) \\
\sin(2\pi B_{\text{s}}\mathbf{x}) \\
\cos(2\pi B_{\text{d}}\mathbf{x})\cos(2 \pi t) \\
\cos(2\pi B_{\text{d}}\mathbf{x})\sin(2 \pi t) \\ 
\sin(2\pi B_{\text{d}}\mathbf{x})\cos(2 \pi t) \\
\sin(2\pi B_{\text{d}}\mathbf{x})\sin(2 \pi t)
\end{bmatrix} \in \mathbb{R}^{2M_{\text{s}} + 4M_{\text{d}}}
\label{eq:fffracmixed}
\end{equation}

Each component of the two matrices \(\subtext{B}{s} \in \mathbb{R}^{\subtext{M}{s} \times 2}\) and \(\subtext{B}{d} \in  \mathbb{R}^{\subtext{M}{d} \times 2}\) are drawn from a normal distribution using the same \(\sigma\) in both cases, as in standard Gaussian Fourier Features. For \(\mathbf{x} \in \mathbf{X}, t \in [0,1]\), the STiFF vector \(\gamma(\mathbf{x}, t; \beta)\) has static components, associated with \(\subtext{B}{s}\), and dynamic components, associated with \(\subtext{B}{d}\). The periodicity of STiFF is enforced by the time-dependent sine and cosine functions with a frequency equals to one in the dynamic components. We note that even though the static and dynamic components are explicitly separated at the input of the MLP, these are sub-sequentially combined by the hidden layers of the network. The length of the STiFF vector is \(L = 2\subtext{M}{s} + 4\subtext{M}{d}\). The percentage \(p_{\text{s}}:= \frac{2\subtext{M}{s}}{L} \cdot 100\) of the STiFF vector that is static, together with the length \(L\) and \(\sigma\) are the hyperparameters of the proposed method. These hyperparameters are summarized under the name \(\beta\).

Therefore, for spatio-temporal coordinates \(\mathbf{x} \in \mathbf{X}, t \in [0,1]\) of cardiac cine MR images, the proposed Intensity Net \(\mathcal{I}_\beta\) with hyperparameters \(\beta\) is defined as:

\begin{equation}
\begin{aligned}
    \mathcal{I}_{\beta}(\mathbf{x}, t; \theta) &:= I_{\beta}(\mathbf{x}, t; \theta)_1 + i I_{\beta}(\mathbf{x}, t; \theta)_2 \in \mathbb{C} \\ 
    \mathrm{where} \,\, I_{\beta}(\mathbf{x}, t; \theta) &:= \mathcal{N}(\gamma(\mathbf{x}, t; \beta); \theta) \in \mathbb{R}^2,
\label{eq:intensity-net}
\end{aligned}
\end{equation}
where \(\mathcal{N}\) is a fully connected MLP with an \(L\)-dimensional input, an output of dimension 2 (real and imaginary), and weights \(\theta\).

The image at a particular cardiac phase \(t\) is obtained by evaluating the Intensity Net on \(t\) and every coordinate in a spatial grid \(G\). For an acquisition of \(N_t\) frames, the \(j\)-th frame is generated with \(t = j/N_t\), for \(j \in [1, N_t]\). The generated image at frame \(t\) is referred to as \(\mathcal{G}_t(\theta):= \{\mathcal{I}_\beta(\mathbf{x},t;\theta)\}_{\mathbf{x}\in G} \in \mathbb{C}^{N_x \times N_y}\), with the full dynamic image referred to as \(\mathcal{G}(\theta) \in \mathbb{C}^{N_x \times N_y \times N_t}\). It is worth noting that, because of the continuous nature of the network, the Intensity Net can also be evaluated on any spatial and temporal coordinates in the domain. The objective is finding \(\theta\) such that \(\mathcal{G}(\theta)\) is a good reconstruction from the acquired data for a given subject. The training strategy used to achieve this is described below. 

\noindent \textbf{Network training.} To reconstruct an image, the network is trained to solve the inverse problem with implicit regularization stated in equation \eqref{eq:implicit-tddip-reg}, where \(\mathcal{G}(\theta)\) is the reconstruction given by Intensity Net. We note that the amount of implicit regularization is controlled by the \(\beta\) hyperparameters of STiFF.


NF-cMRI is proposed here for radially acquired data \(\mathbf{b}\). For reconstruction of radial acquisition trajectories, typically the NUFFT is considered in the forward model \cite{fengGoldenAngleRadialMRI2022}. Here, we propose a simpler implementation that combines the Fourier Slice Theorem with the continuous nature of our network. A diagram of the training process can be seen in Figure \ref{fig:training}.

Specifically, consider a radial acquisition from coil \(c \in [1,N_c]\) at cardiac phase \(t \in [0,1]\) with direction of the unit vector \(\boldsymbol{\xi}\in\mathbb{R}^2\). We call this measurement \(\mathbf{b}_{\xi, c}\) acquired at frequencies along the direction \(\boldsymbol{\xi}\), defined as $\mathbf{k}_r = \{k_r^i\boldsymbol{\xi}\}_{i \in [1,N_k]}$ with $k_r^i = k_{min} + i/N_k(k_{max} - k_{min})$. Using a known coil sensitivity $S_c$, a sensitivity-weighted reconstruction can be generated with the Intensity Net as \(\mathcal{G}_t(\theta) \odot S_c\), where \(\odot\) is an element-wise product. To evaluate the loss function, the Fourier transform of the proposed reconstruction needs to be computed at frequencies $\mathbf{k}_r$ along the direction \(\boldsymbol{\xi}\). This can be done using the Fourier Slice Theorem, which states that the 2D Fourier transform along one line can be computed by first taking the Radon transform and then a 1D Fourier transform on the result.
The Fourier Slice Theorem states that formula \eqref{eq:fourier-slice} holds for any square-integrable function \(f\) that vanishes outside some bounded set, which is the case for a full FOV acquisition:
\begin{equation}
\mathcal{F}_1 \big[\mathcal{R}_{\boldsymbol{\xi}}[f] \big](k_r^i) = \mathcal{F}_2 [f](k_r^i \boldsymbol{\xi}).\label{eq:fourier-slice}
\end{equation}

\(\mathcal{F}_{d}\) stands for the \(d\)-dimensional unnormalized Fourier Transform, and \(\mathcal{R}_\xi[f]\) is the Radon Transform, which is calculated by integrating \(f\) along lines with direction \(\boldsymbol{\xi}\):

\begin{equation}
\mathcal{R}_{\boldsymbol{\xi}}[f](s) := \int_\mathbb{R} f(s\boldsymbol{\xi}+\tau\boldsymbol{\xi}^\perp) d\tau, \quad s\in\mathbb{R},\label{eq:rad-def}
\end{equation}
where  $\boldsymbol{\xi}^\perp$ is the counterclockwise rotation of $\boldsymbol{\xi}$ by 90 degrees.

Then, we can compute the 2-dimensional Fourier Transform of the reconstructed image at a frequency along the acquired direction \(\boldsymbol{\xi}\) as:
\begin{equation}
     g^{i}_{\boldsymbol{\xi},c}(\theta) :=  \mathcal{F}_1 \Big[ \mathcal{R}_{\boldsymbol{\xi}} \big[ \mathcal{G}_{t}(\theta) \odot S_c \big] \Big](k_r^i).
    \label{eq:imggen-fourier}
\end{equation}
We note that we only require one evaluation of the image $\mathcal{G}_{t}(\theta)$ from Intensity Net to compute the Fourier transform along the radial trajectory. This is achieved by computing first the Radon transform and then computing a 1D FFT on the result.  The continuous nature of the Intensity Net allows calculating the Radon transform \(\mathcal{R}_{\boldsymbol{\xi}}\big[ \mathcal{G}_{t}(\theta)\big]\) easily; just evaluating \(\mathcal{I}_\beta\) in a rotation of the grid \(G\) in the angle of \(\boldsymbol{\xi}\), and summation over one variable. This eliminates the need of interpolating to obtain a rotated image. 

The loss associated for all coils $c$ and one radial trajectory is given by:
\begin{equation}
    \mathcal{L}_{\boldsymbol{\xi}}(\theta) := \frac{1}{N_k N_c}\sum_{i = 1}^{N_k} \sum_{c = 1}^{N_c} \Big( \big(1 + |k_r^i|\big)\big(g^{i}_{\boldsymbol{\xi},c}(\theta) - b^{i}_{\boldsymbol{\xi},c} \big) \Big)^2.
    \label{eq:spoke-loss}
\end{equation}

The weights \((1+|k_r|)\) are included such that the loss function becomes equivalent to the mean square distance in the image space\cite{nattererMathematicsComputerizedTomography2001}. The complete loss function is defined as:
\begin{equation}
    \mathcal{L}(\theta) := \frac{1}{N_{\boldsymbol{\xi}}} \sum_{\boldsymbol{\xi} \in \Xi} \mathcal{L}_{\boldsymbol{\xi}}(\theta),
\end{equation}
where $\Xi$ is the set of all radial measured directions $\boldsymbol{\xi}$. For efficient training, we can select a subset of radial directions $\boldsymbol{\xi}$ for mini-batch.

\section{Experimental Setup} \label{experiments}

The proposed method is evaluated on a publicly available clinical dataset \cite{el-rewaidyReplicationDataMultiDomain2020}. This dataset contains radial bSSFP cine data collected from 108 subjects (101 patients and 7 healthy subjects) with a 3T MRI scanner (MAGNETOM Vida, Siemmens). Data were acquired with retrospective ECG-triggering, 25 cardiac phases, 196 radial spokes (partial echo), breath-hold ~14 heartbeats, TR/TE =3.06/1.4ms, FOV = 380 x 380 mm$^2$, in-plane resolution = 1.8x1.8 mm$^2$, slice thickness = 8 mm, flip angle = 48°, channels = 16 ± 1. The acquisition is nearly-fully-sampled (acceleration 1.06x with respect to Cartesian fully sampled). Undersampling is performed by selecting $n = 8$ or $4$ spokes per frame of the total of $196$ radial spokes (acceleration factors of 26x, and 52x respectively). The spokes are selected in a "pseudo-golden-angle" fashion: a series of desired angles with golden-angle progression are chosen and binned into $n$ angles per frame, similar to the undersampling performed by \cite{el-rewaidyMultidomainConvolutionalNeural2021}.

A ground truth fully sampled reconstruction is calculated using GRASP \cite{fengGoldenangleRadialSparse2014}, as implemented in BART \cite{blumenthalMrireconBartVersion2022}. The hyperparameters $\lambda = 0.001, \mathscr{L}= 0.0005$ and 100 iterations were chosen by visual inspection on a single representative patient and used for the remaining ground truth reconstructions. Coil sensitivities are estimated from the data itself using ESPIRiT \cite{ueckerESPIRiTEigenvalueApproach2014}, also using BART. 

Undersampled reconstruction was performed with the proposed approach and with GRASP for comparison purposes. The  $\lambda$ and $\mathscr{L}$ hyperparameters of GRASP were optimized for the same single representative patient, using grid search in $\lambda \in {1 \times 10^{-6}, \dots, 1 \times 10^1}$  and $\mathscr{L} \in {5 \times 10^{-4}, \dots, 5 \times 10^1}$. This optimization is performed independently for each acceleration factor. The \textit{Structural Similarity Index Metric} (SSIM) \cite{wangImageQualityAssessment2004} from the scikit-image package \cite{waltScikitimageImageProcessing2014}, with default parameters, was used to choose the best $\lambda$ and $\mathscr{L}$. The 3D SSIM (2D + time) was calculated taking the ground truth images as reference, and comparing in a cropped area around the heart. The selected values were $\lambda = 0.001, \mathscr{L}= 5.0$ for 26x and $\lambda = 0.01, \mathscr{L}= 5.0$ for 52x. All the reconstructions are done with 100 iterations. Reconstruction time was $\approx$ 5 min.

The hyperparameters $\beta = (p_{\text{s}}, L, \sigma)$ of the proposed STiFF preprocessing are optimized for the same single representative patient in a similar fashion to the GRASP optimization; a grid search is performed for $p_{\text{s}} \in \{67, 80, 90\}, L \in \{400, 600, 800\}$ and $\sigma \in \{4.5, 5.5, 6.5, 7.5\}$, choosing the combinations which result in the best 3D SSIM when comparing to the ground truth reconstruction. Once the hyperparameters for both GRASP and STiFF are selected, the same values are used to generate undersampled reconstructions for the first six equally-sampled patients of the dataset.

All the experiments were run in a machine with AMD Ryzen 5 5600X 6-Core processor and graphics card Nvidia GeForce RTX 3080 with 10GB. Batching is done considering all the available radial spokes for the undersampled acquisition, with batch size of 2. The MLP have 3 inner layers of 250 nodes and uses ReLU as activation function. The network weights are optimized using ADAM \cite{kingmaAdamMethodStochastic2014} with learning rate $7.5 \times 10^{-4}$ and $10\,000$ iterations, taking $\approx$ 10 min. The implementation is done in JAX \cite{bradburyJAXComposableTransformations2018} and is available in the GitHub repository \url{https://github.com/fsahli/NF-cMRI}.


\section{Results} \label{results}

\begin{figure}[!t]
    \centerline{\includegraphics[width=0.8\textwidth]{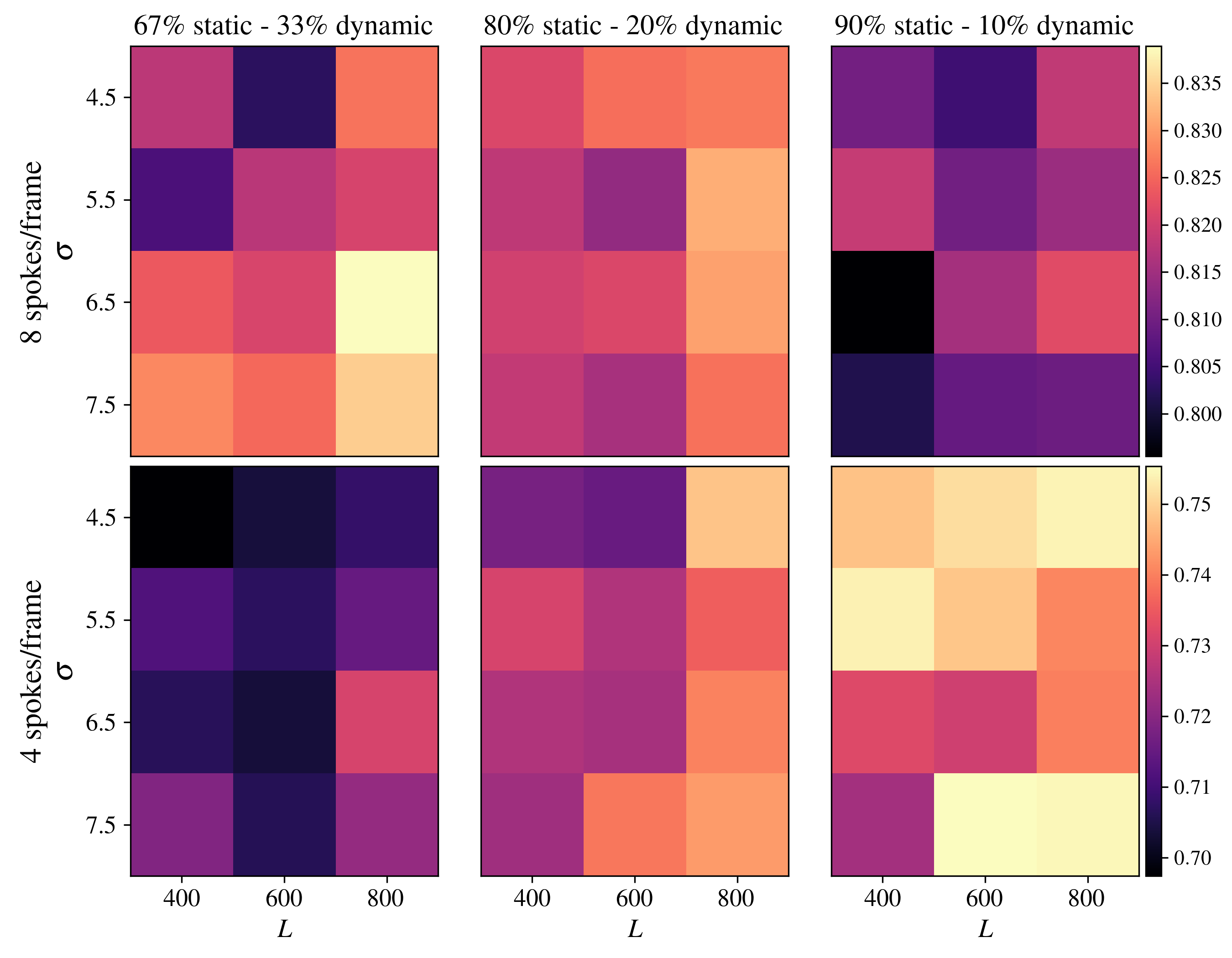}}
    \caption{SSIM 3D based Hyperparameter optimization for STiFF, for different acceleration factors. Each row corresponds to an acceleration factor (26x and 52x); color scales are comparable along the row. The columns group different static fractions for the dynamic components \(\subtext{p}{s} = \perc{67}, \perc{80}\) and \(\perc{90}\).}
    \label{fig:hyper-frac-ss}
\end{figure}

\noindent\textbf{STiFF Hyperparameter optimization.} SSIM-based hyperparemeter optimization results for the proposed STiFF are shown in Figure \ref{fig:hyper-frac-ss} for acceleration factors of 26x (8 spokes/frame) and 52x (4 spokes/frame), for different static fractions for the dynamic components \(\subtext{p}{s} = \perc{67}, \perc{80}\) and \(\perc{90}\), and for different \(L\). Results show that hyperparameter selection plays a big role in the quality of the image generated with the proposed STiFF (as is the case for Gaussian Fourier Features). Small \(\sigma\) can lead to blurring whereas large \(\sigma\) can produce images with spatial-oscillations artifacts. On the other hand, a higher mapping size improves the learning process and is only limited by the computation burden. Figure 3 shows a strong dependency on the acceleration factor for the ideal fraction of static features \(\subtext{p}{s}\) in the proposed STiFF method. This parameter acts as a regularizer; a higher \(\subtext{p}{s}\) results in  stronger temporal regularization and can introduce temporal blurring. Optimal parameters depend on the acceleration factor. For 4 spokes/frame, the best reconstructions correspond to \(\subtext{p}{s}=\perc{67}\), whereas for 4 spokes/frame best results are obtained for \(\subtext{p}{s}=\perc{90}\). Based on the metrics showed in Figure \ref{fig:hyper-frac-ss}, for all subsequent reconstructions we choose \(\sigma=6.5, L=800\) and \(\subtext{p}{s}=\perc{67}\) for reconstructions using 8 spokes/frame, and \(\sigma=7.5, L=600\) and \(\subtext{p}{s}=\perc{90}\) for 4 spokes/frame.

\begin{figure}[!t]
    \centerline{\includegraphics[width=0.75\textwidth]{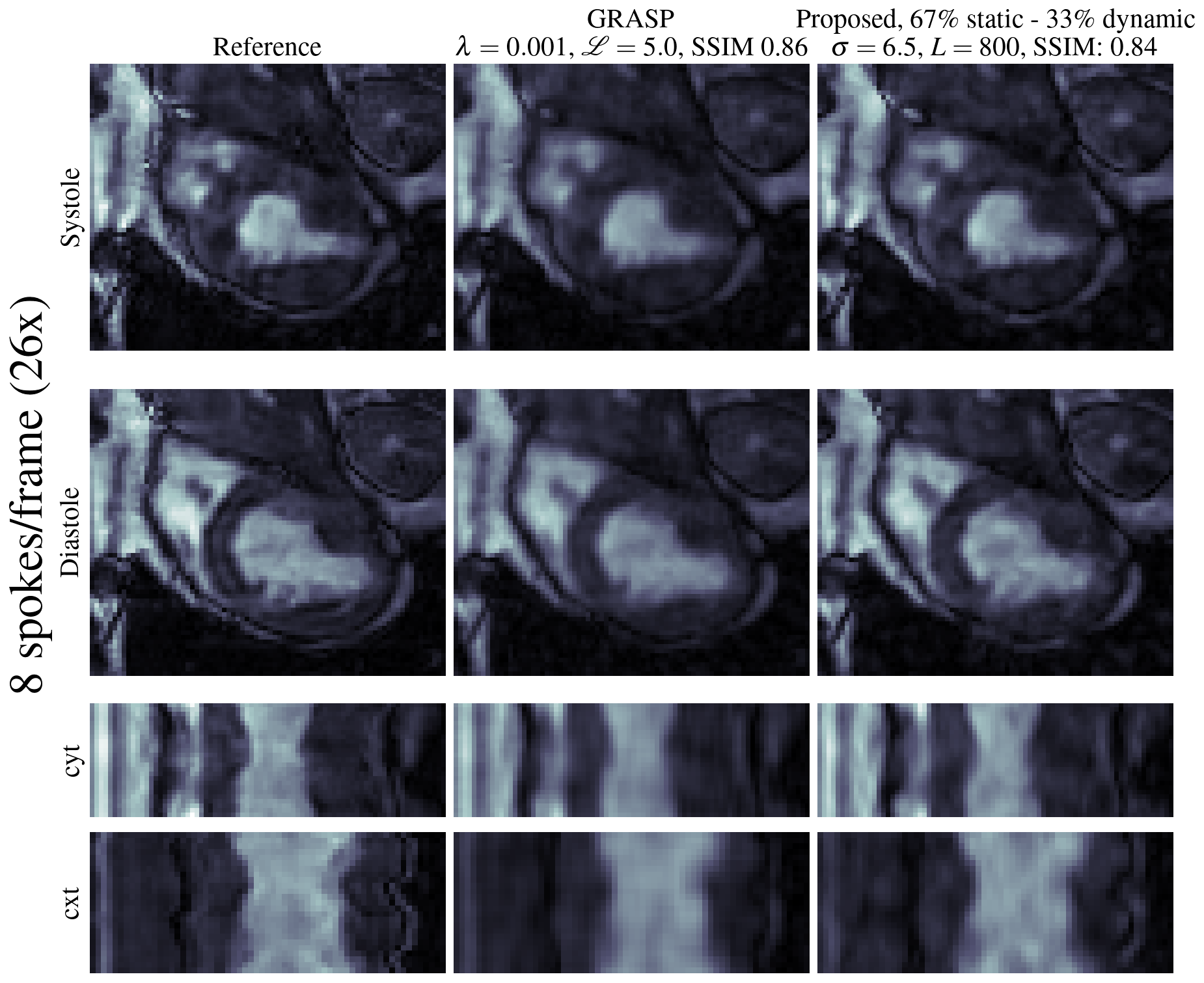}}
    \centerline{\includegraphics[width=0.75\textwidth]{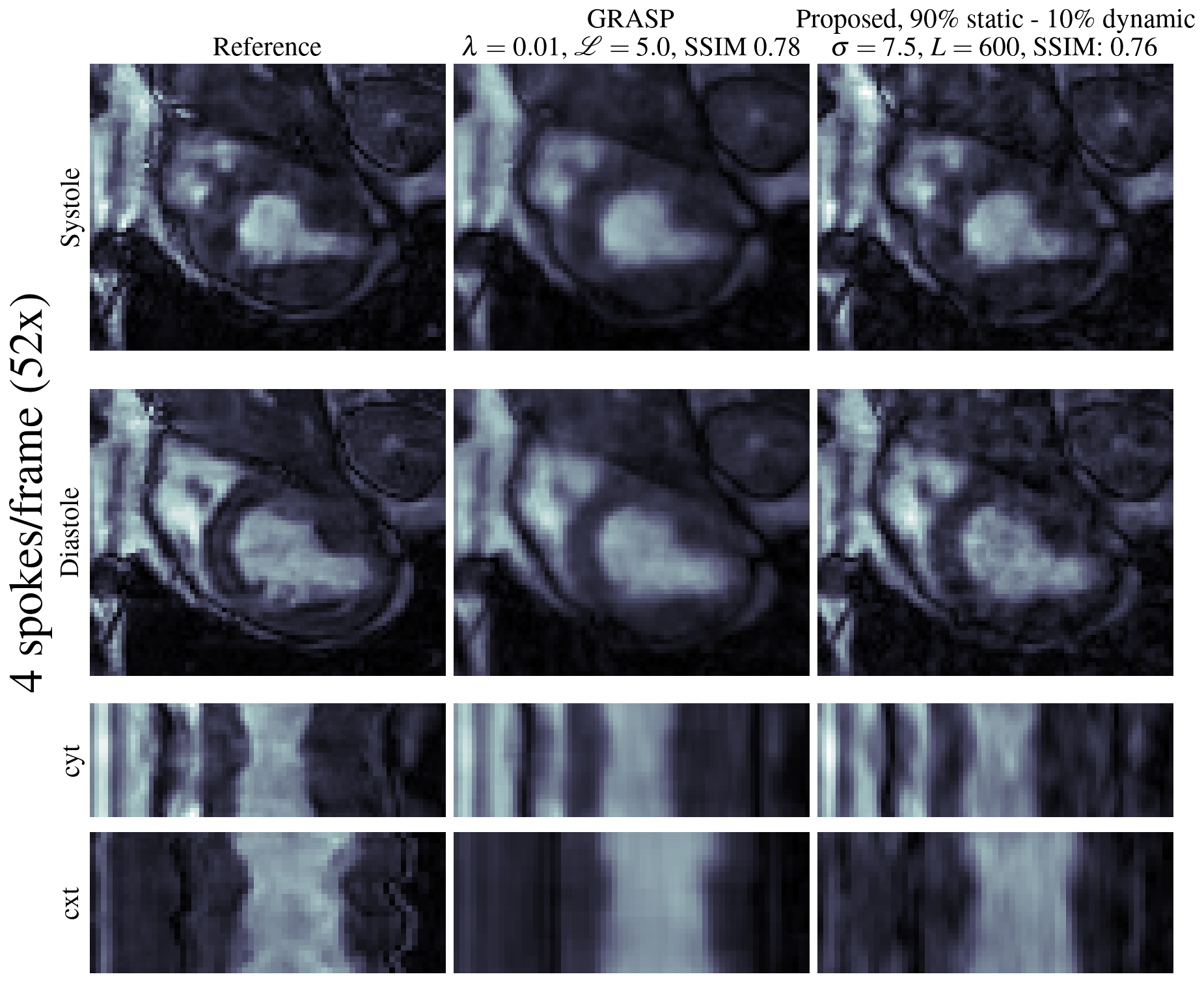}}
    \caption{Reconstruction using the proposed method in a representative patient, compared against GRASP and reference images, for 8 and 4 spokes/frame. The following views are shown: the cropped area around the heart used for the 3D SSIM calculation shown for systole and diastole, and the temporal evolution of horizontal {\textbf{cyt}} and vertical {\textbf{cxt}} cuts along the center.}
    \label{fig:best-against-grasp} 
\end{figure}

\noindent\textbf{Comparison against GRASP.} Reconstruction results for the proposed NF-cMRI for 4 and 8 spokes/frame are shown in Figure \ref{fig:best-against-grasp} for systole, diastole and for horizontal and vertical temporal evolution. Results corresponds to the representative case used for hyperparemeter optimization. Corresponding fully sampled images and GRASP reconstruction are also included in Figure \ref{fig:best-against-grasp} for comparison purposes. For both acceleration factors, the proposed method achieved comparable spatial and better temporal visual image quality in comparison to GRASP. Slightly lower SSIM values were obtained with the proposed approach in comparison to GRASP.



\begin{figure*}[!t]
    \centerline{\includegraphics[width=\textwidth]{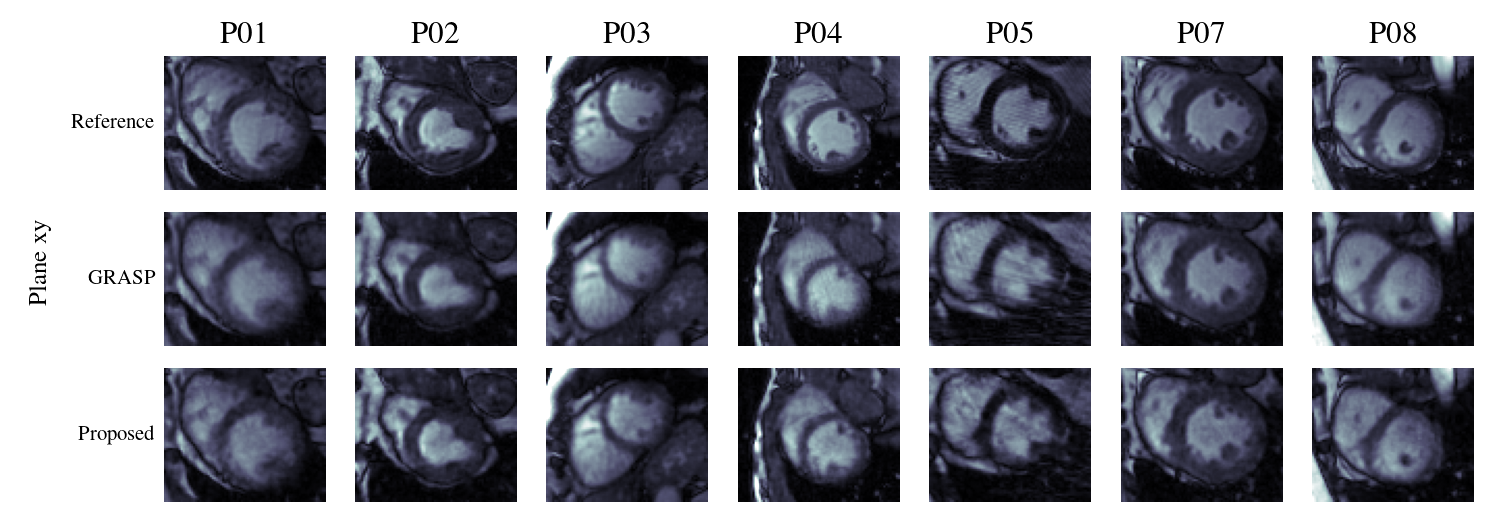}}
    \centerline{\includegraphics[width=\textwidth]{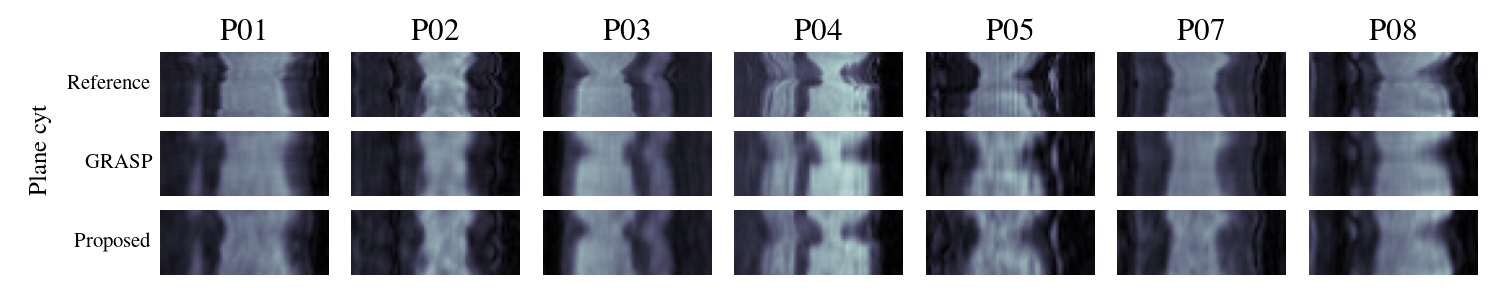}}
    \caption{Reconstruction results for the proposed NF-cMRI for 8 spokes/frame for all patients in comparison to fully sampled reference (first row) and GRASP reconstruction (second row). Hyper-parameter optimization was done in patient P02.}
    \label{fig:all-patients}
\end{figure*}

\begin{figure*}[!t]
    \centerline{\includegraphics[width=\textwidth]{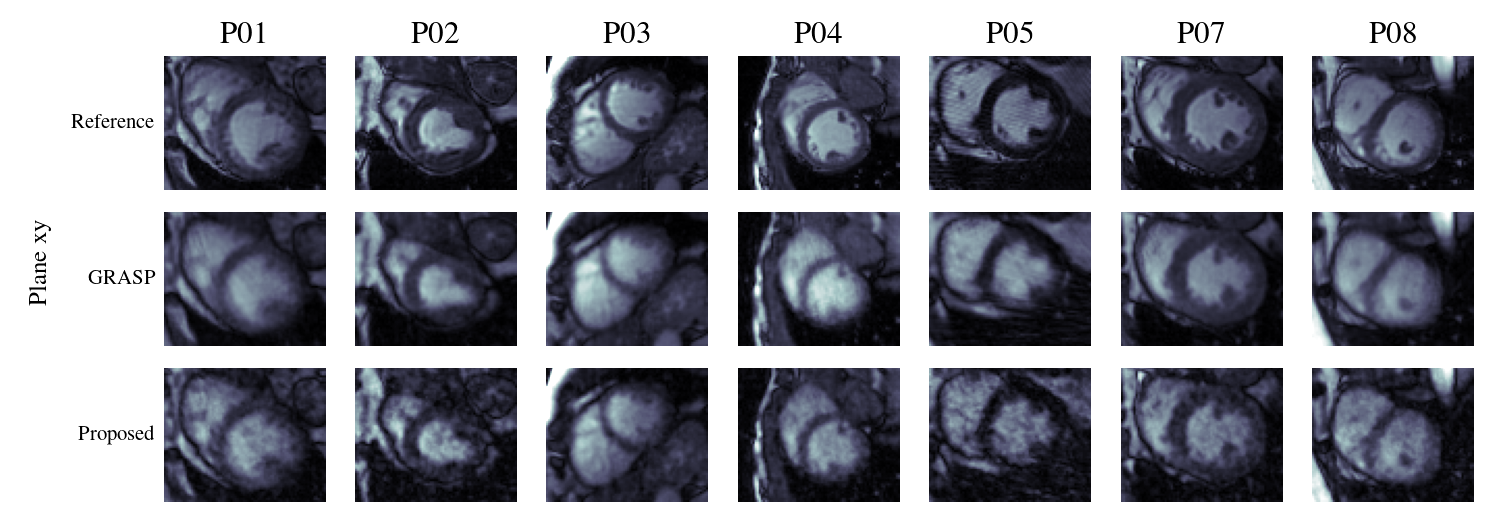}}
    \centerline{\includegraphics[width=\textwidth]{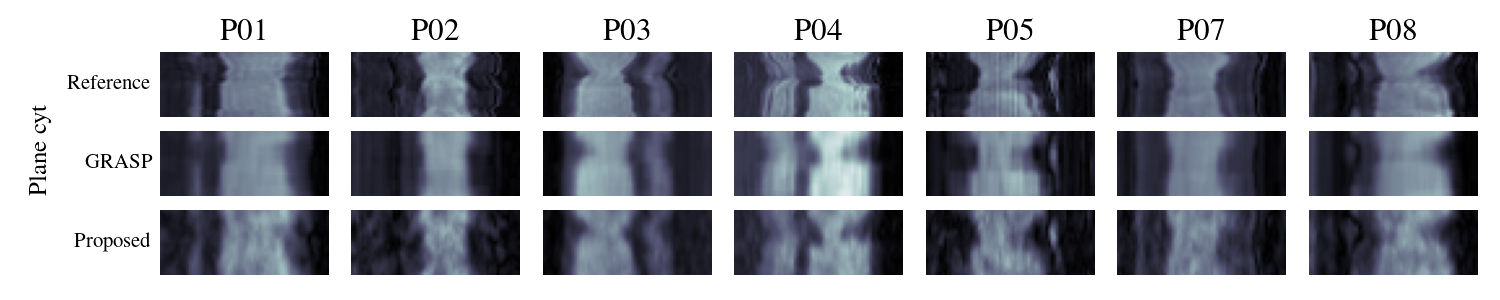}}
    \caption{Reconstruction results for the proposed NF-cMRI for 4 spokes/frame for all patients in comparison to fully sampled reference (first row) and GRASP reconstruction (second row). Hyper-parameter optimization was done in patient P02.}
    \label{fig:all-patients4}
\end{figure*}

\noindent\textbf{Hyperparameter generalization.} Reconstruction results for the proposed NF-cMRI for 8 spokes/frame are shown in Figure \ref{fig:all-patients} for all patients, and in Figure \ref{fig:all-patients4} for 4 spokes/frame. Corresponding fully sampled images and GRASP reconstruction are also included these figures for comparison purposes. The proposed method achieved comparable or better visual image quality in comparison to GRASP. Chosen hyperparameters generalize well to the remaining patients.  



\section{Discussion} \label{discussion}


We have presented NF-cMRI, an unsupervised deep learning reconstruction method for multi-coil golden radial cardiac cine MRI data. The proposed approach was evaluated for undersampling factors of 26x and 52x, achieving comparable results to state-of-the-art GRASP reconstruction. NF-cMRI creates a continuous representation of cine MRI using neural fields, enhanced with a spatio-temporal encoding. The proposed method is unsupervised and can be trained on an undersampled acquisition from a single subject, not relying on large databases.

The main contributions of this work are
(1) We proposed STiFF, an extension to  for working with nearly-periodic dynamic images. The STiFF hyperparameters allow to adequately control the strength of the regularization in the spatial and temporal dimensions.
(2) We use the Fourier-Slice Theorem to implement a simple version of the Fourier Transform, which allowed processing the multi-coil radial acquisitions individually, without the need of the NUFFT. 
(3) Our work shows the feasibility of using implicit neural representations for \textit{in-vivo} undersampled data reconstruction. We achieved comparable metrics and improved temporal image quality against GRASP in highly accelerated acquisitions.


Results show that, similar to GRASP, the performance of the method depends on a good choice of hyperparameters. In particular, the static fraction \(\subtext{p}{s}\) election must take into account the acceleration factor. A stronger regularization, given by a higher value of \(\subtext{p}{s}\), is necessary to reduce artifacts in reconstructions with higher accelerations, however at the expense of introducing some temporal blurring. Hyperparameters seem to generalize well between different acquisitions, although they should depend on the dynamic image resolution. 

The obtained results are comparable to GRASP in terms of SSIM; with slightly higher SSIM values for GRASP. The proposed method obtains visually better results and presents perceivable higher temporal fidelity, as shown by the spatio-temporal cuts in Figures 4-6. 


The proposed approach has some limitations. The method has some difficulties identifying regions of temporal variation and restricting temporal changes to such regions. In particular, reducing temporal regularization to improve motion recovery may introduced artifacts that oscillate in time in motionless regions. This was particularly noticeable during our preliminary experimentation with higher acceleration factors and lower \(\subtext{p}{s}\) values. The improved temporal fidelity was associated with a lower SSIM, as this metric tends to favor smoother images \cite{muckleyResults2020FastMRI2021}. In the future, we plan to explore other spatio-temporal encodings, such as the wavelet transform, which may be better to spatially localize the temporal variations in the image \cite{saragadamWIREWaveletImplicit2023}. 

The proposed approach requires training times of about 10 minutes, which is comparable to the GRASP reconstruction times. This is not of primary importance with supervised methods, as they are usually trained just once, and the inference time is considerably faster. On the contrary, NF-cMRI needs to be retrained with every new acquisition. This is a common problem of unsupervised methods, as is the case for TD-DIP \cite{yooTimeDependentDeepImage2021}. Reducing training time is also important and an active area of research in neural fields. For example,\cite{mullerInstantNeuralGraphics2022} proposed the use of parametric encodings with optimizable parameters (hash encoding) instead of  to reduce training times. However, this approach requires explicit additional regularization (\cite{fengSpatiotemporalImplicitNeural2023}, \cite{kunzImplicitNeuralNetworks2023}). And using transfer learning from reconstruction of other patients may suffer the same limitations of the supervised approaches, where some bias may be introduced in the process.

By construction, STiFF works with nearly-periodic cardiac cycles. This is a reasonable assumption for most cardiac cine acquisitions and a limitation of most current methods. However, this assumption is not satisfied for patients with arrhythmias. It is not difficult to extend STiFF to include multiple different frequencies in the temporal direction and attempt real time reconstruction of an arrhythmic heart.

In this work, we focused on radial acquisition, for which we developed an efficient method to evaluate the loss function. We note that our method is specially efficient when a low number of radial acquisitions $N_{\boldsymbol{\xi}}$ is used for the reconstruction. The computational complexity of our method is $O(N_{\boldsymbol{\xi}}N_k^2)$, while the NUFFT has complexity $O(N_k^2\log(N_k^2))$, which is comparable for low values of $N_{\boldsymbol{\xi}}$. Applying this method to Cartesian acquisitions only requires a modification of the evaluation of the loss function, which should be simplified by the direct use of the FFT.

In summary, we have developed a method to reconstruct cardiac cine MRI using neural fields. This technique could reduce scan times that may lead to more and better access of patients to this technology.










\section*{Acknowledgments}
This work was supported by the following grants: (1) Millennium Nucleus for Applied Control and Inverse Problems ACIP NCN19\_161, (2) Millennium Institute for Intelligent Healthcare Engineering iHEALTH ICN2021\_004. (3) Anid Basal FB210005.

\bibliographystyle{unsrt}  
\bibliography{better-biblioteca}

\begin{thebibliography}{10}

\bibitem{mojibianChapterCardiacMagnetic2018}
Hamid Mojibian and Hamidreza Pouraliakbar.
\newblock Chapter 8 - {{Cardiac Magnetic Resonance Imaging}}.
\newblock In Majid Maleki, Azin Alizadehasl, and Majid Haghjoo, editors, {\em
  Practical {{Cardiology}}}, pages 159--166. {Elsevier}, January 2018.

\bibitem{donevaMathematicalModelsMagnetic2020}
Mariya Doneva.
\newblock Mathematical {{Models}} for {{Magnetic Resonance Imaging
  Reconstruction}}: {{An Overview}} of the {{Approaches}}, {{Problems}}, and
  {{Future Research Areas}}.
\newblock {\em IEEE Signal Processing Magazine}, 37(1):24--32, January 2020.

\bibitem{knollDeepLearningMethodsParallel2020}
Florian Knoll, Kerstin Hammernik, Chi Zhang, Steen Moeller, Thomas Pock,
  Daniel~K. Sodickson, and Mehmet Akcakaya.
\newblock Deep-{{Learning Methods}} for {{Parallel Magnetic Resonance Imaging
  Reconstruction}}: {{A Survey}} of the {{Current Approaches}}, {{Trends}}, and
  {{Issues}}.
\newblock {\em IEEE Signal Processing Magazine}, 37(1):128--140, January 2020.

\bibitem{pruessmannSENSESensitivityEncoding1999}
K.~P. Pruessmann, M.~Weiger, M.~B. Scheidegger, and P.~Boesiger.
\newblock {{SENSE}}: Sensitivity encoding for fast {{MRI}}.
\newblock {\em Magnetic Resonance in Medicine}, 42(5):952--962, November 1999.

\bibitem{griswoldGeneralizedAutocalibratingPartially2002}
Mark~A. Griswold, Peter~M. Jakob, Robin~M. Heidemann, Mathias Nittka, Vladimir
  Jellus, Jianmin Wang, Berthold Kiefer, and Axel Haase.
\newblock Generalized autocalibrating partially parallel acquisitions
  ({{GRAPPA}}).
\newblock {\em Magnetic Resonance in Medicine}, 47(6):1202--1210, June 2002.

\bibitem{lustigCompressedSensingMRI2008}
Michael Lustig, David~L. Donoho, Juan~M. Santos, and John~M. Pauly.
\newblock Compressed {{Sensing MRI}}.
\newblock {\em IEEE Signal Processing Magazine}, 25(2):72--82, March 2008.

\bibitem{menchon-laraReconstructionTechniquesCardiac2019}
Rosa-Mar{\'i}a {Mench{\'o}n-Lara}, Federico {Simmross-Wattenberg}, Pablo
  {Casaseca-de-la-Higuera}, Marcos {Mart{\'i}n-Fern{\'a}ndez}, and Carlos
  {Alberola-L{\'o}pez}.
\newblock Reconstruction techniques for cardiac cine {{MRI}}.
\newblock {\em Insights into Imaging}, 10:100, September 2019.

\bibitem{fengGoldenangleRadialSparse2014}
Li~Feng, Robert Grimm, Kai~Tobias Block, Hersh Chandarana, Sungheon Kim, Jian
  Xu, Leon Axel, Daniel~K. Sodickson, and Ricardo Otazo.
\newblock Golden-angle radial sparse parallel {{MRI}}: Combination of
  compressed sensing, parallel imaging, and golden-angle radial sampling for
  fast and flexible dynamic volumetric {{MRI}}.
\newblock {\em Magnetic Resonance in Medicine}, 72(3):707--717, September 2014.

\bibitem{koflerSpatioTemporalDeepLearningBased2020a}
Andreas Kofler, Marc Dewey, Tobias Schaeffter, Christian Wald, and Christoph
  Kolbitsch.
\newblock Spatio-{{Temporal Deep Learning-Based Undersampling Artefact
  Reduction}} for {{2D Radial Cine MRI With Limited Training Data}}.
\newblock {\em IEEE Transactions on Medical Imaging}, 39(3):703--717, March
  2020.

\bibitem{zhuImageReconstructionDomaintransform2018}
Bo~Zhu, Jeremiah~Z. Liu, Stephen~F. Cauley, Bruce~R. Rosen, and Matthew~S.
  Rosen.
\newblock Image reconstruction by domain-transform manifold learning.
\newblock {\em Nature}, 555(7697):487--492, March 2018.

\bibitem{akcakayaScanspecificRobustArtificialneuralnetworks2019}
Mehmet Ak{\c c}akaya, Steen Moeller, Sebastian Weing{\"a}rtner, and K{\^a}mil
  U{\u g}urbil.
\newblock Scan-specific robust artificial-neural-networks for k-space
  interpolation ({{RAKI}}) reconstruction: {{Database-free}} deep learning for
  fast imaging.
\newblock {\em Magnetic Resonance in Medicine}, 81(1):439--453, 2019.

\bibitem{schlemperDeepCascadeConvolutional2018}
Jo~Schlemper, Jose Caballero, Joseph~V. Hajnal, Anthony~N. Price, and Daniel
  Rueckert.
\newblock A {{Deep Cascade}} of {{Convolutional Neural Networks}} for {{Dynamic
  MR Image Reconstruction}}.
\newblock {\em IEEE Transactions on Medical Imaging}, 37(2):491--503, February
  2018.

\bibitem{hammernikPhysicsDrivenDeepLearning2023}
Kerstin Hammernik, Thomas K{\"u}stner, Burhaneddin Yaman, Zhengnan Huang,
  Daniel Rueckert, Florian Knoll, and Mehmet Ak{\c c}akaya.
\newblock Physics-{{Driven Deep Learning}} for {{Computational Magnetic
  Resonance Imaging}}: {{Combining}} physics and machine learning for improved
  medical imaging.
\newblock {\em IEEE Signal Processing Magazine}, 40(1):98--114, January 2023.

\bibitem{hammernikSystematicEvaluationIterative2021}
Kerstin Hammernik, Jo~Schlemper, Chen Qin, Jinming Duan, Ronald~M. Summers, and
  Daniel Rueckert.
\newblock Systematic evaluation of iterative deep neural networks for fast
  parallel {{MRI}} reconstruction with sensitivity-weighted coil combination.
\newblock {\em Magnetic Resonance in Medicine}, 86(4):1859--1872, 2021.

\bibitem{muckleyResults2020FastMRI2021}
Matthew~J. Muckley, Bruno Riemenschneider, Alireza Radmanesh, Sunwoo Kim, Geunu
  Jeong, Jingyu Ko, Yohan Jun, Hyungseob Shin, Dosik Hwang, Mahmoud Mostapha,
  Simon Arberet, Dominik Nickel, Zaccharie Ramzi, Philippe Ciuciu, Jean-Luc
  Starck, Jonas Teuwen, Dimitrios Karkalousos, Chaoping Zhang, Anuroop Sriram,
  Zhengnan Huang, Nafissa Yakubova, Yvonne~W. Lui, and Florian Knoll.
\newblock Results of the 2020 {{fastMRI Challenge}} for {{Machine Learning MR
  Image Reconstruction}}.
\newblock {\em IEEE transactions on medical imaging}, 40(9):2306--2317,
  September 2021.

\bibitem{antunInstabilitiesDeepLearning2020}
Vegard Antun, Francesco Renna, Clarice Poon, Ben Adcock, and Anders~C. Hansen.
\newblock On instabilities of deep learning in image reconstruction and the
  potential costs of {{AI}}.
\newblock {\em Proceedings of the National Academy of Sciences},
  117(48):30088--30095, December 2020.

\bibitem{ulyanovDeepImagePrior2020}
Dmitry Ulyanov, Andrea Vedaldi, and Victor Lempitsky.
\newblock Deep {{Image Prior}}.
\newblock {\em International Journal of Computer Vision}, 128(7):1867--1888,
  July 2020.

\bibitem{qayyumUntrainedNeuralNetwork2023}
Adnan Qayyum, Inaam Ilahi, Fahad Shamshad, Farid Boussaid, Mohammed Bennamoun,
  and Junaid Qadir.
\newblock Untrained {{Neural Network Priors}} for {{Inverse Imaging Problems}}:
  {{A Survey}}.
\newblock {\em IEEE Transactions on Pattern Analysis and Machine Intelligence},
  45(5):6511--6536, May 2023.

\bibitem{zhaoReferenceDrivenCompressedSensing2020}
Di~Zhao, Feng Zhao, and Yongjin Gan.
\newblock Reference-{{Driven Compressed Sensing MR Image Reconstruction Using
  Deep Convolutional Neural Networks}} without {{Pre-Training}}.
\newblock {\em Sensors}, 20(1):308, January 2020.

\bibitem{aggarwalENSUREGeneralApproach2023}
Hemant~Kumar Aggarwal, Aniket Pramanik, Maneesh John, and Mathews Jacob.
\newblock {{ENSURE}}: {{A General Approach}} for {{Unsupervised Training}} of
  {{Deep Image Reconstruction Algorithms}}.
\newblock {\em IEEE Transactions on Medical Imaging}, 42(4):1133--1144, April
  2023.

\bibitem{yooTimeDependentDeepImage2021}
Jaejun Yoo, Kyong~Hwan Jin, Harshit Gupta, Jerome Yerly, Matthias Stuber, and
  Michael Unser.
\newblock Time-{{Dependent Deep Image Prior}} for {{Dynamic MRI}}, January
  2021.

\bibitem{zouDynamicImagingUsing2021a}
Qing Zou, Abdul~Haseeb Ahmed, Prashant Nagpal, Stanley Kruger, and Mathews
  Jacob.
\newblock Dynamic {{Imaging Using}} a {{Deep Generative SToRM}} ({{Gen-SToRM}})
  {{Model}}.
\newblock {\em IEEE Transactions on Medical Imaging}, 40(11):3102--3112,
  November 2021.

\bibitem{xieNeuralFieldsVisual2022}
Yiheng Xie, Towaki Takikawa, Shunsuke Saito, Or~Litany, Shiqin Yan, Numair
  Khan, Federico Tombari, James Tompkin, Vincent Sitzmann, and Srinath Sridhar.
\newblock Neural {{Fields}} in {{Visual Computing}} and {{Beyond}}.
\newblock {\em Computer Graphics Forum}, 41(2):641--676, 2022.

\bibitem{rahamanSpectralBiasNeural2019}
Nasim Rahaman, Aristide Baratin, Devansh Arpit, Felix Draxler, Min Lin, Fred
  Hamprecht, Yoshua Bengio, and Aaron Courville.
\newblock On the {{Spectral Bias}} of {{Neural Networks}}.
\newblock In {\em Proceedings of the 36th {{International Conference}} on
  {{Machine Learning}}}, pages 5301--5310. {PMLR}, May 2019.

\bibitem{basriConvergenceRateNeural2019}
R.~Basri, D.~Jacobs, Y.~Kasten, and S.~Kritchman.
\newblock The {{Convergence Rate}} of {{Neural Networks}} for {{Learned
  Functions}} of {{Different Frequencies}}.
\newblock In {\em Neural {{Information Processing Systems}}}, June 2019.

\bibitem{sitzmannImplicitNeuralRepresentations2020}
Vincent Sitzmann, Julien N.~P. Martel, Alexander~W. Bergman, David~B. Lindell,
  and Gordon Wetzstein.
\newblock Implicit {{Neural Representations}} with {{Periodic Activation
  Functions}}, June 2020.

\bibitem{lindellBaconBandlimitedCoordinate2022}
David~B. Lindell, Dave Van~Veen, Jeong~Joon Park, and Gordon Wetzstein.
\newblock Bacon: {{Band-limited Coordinate Networks}} for {{Multiscale Scene
  Representation}}.
\newblock In {\em 2022 {{IEEE}}/{{CVF Conference}} on {{Computer Vision}} and
  {{Pattern Recognition}} ({{CVPR}})}, pages 16231--16241, {New Orleans, LA,
  USA}, June 2022. {IEEE}.

\bibitem{mullerInstantNeuralGraphics2022}
Thomas M{\"u}ller, Alex Evans, Christoph Schied, and Alexander Keller.
\newblock Instant neural graphics primitives with a multiresolution hash
  encoding.
\newblock {\em ACM Transactions on Graphics}, 41(4):1--15, July 2022.

\bibitem{tancikFourierFeaturesLet2020a}
Matthew Tancik, Pratul Srinivasan, Ben Mildenhall, Sara {Fridovich-Keil},
  Nithin Raghavan, Utkarsh Singhal, Ravi Ramamoorthi, Jonathan Barron, and Ren
  Ng.
\newblock Fourier {{Features Let Networks Learn High Frequency Functions}} in
  {{Low Dimensional Domains}}.
\newblock In {\em Advances in {{Neural Information Processing Systems}}},
  volume~33, pages 7537--7547. {Curran Associates, Inc.}, 2020.

\bibitem{shenNeRPImplicitNeural2022}
Liyue Shen, John Pauly, and Lei Xing.
\newblock {{NeRP}}: {{Implicit Neural Representation Learning With Prior
  Embedding}} for {{Sparsely Sampled Image Reconstruction}}.
\newblock {\em IEEE Transactions on Neural Networks and Learning Systems},
  pages 1--13, 2022.

\bibitem{xuNeSVoRImplicitNeural2023}
Junshen Xu, Daniel Moyer, Borjan Gagoski, Juan~Eugenio Iglesias,
  P.~Ellen~Grant, Polina Golland, and Elfar Adalsteinsson.
\newblock {{NeSVoR}}: {{Implicit Neural Representation}} for {{Slice-to-Volume
  Reconstruction}} in {{MRI}}.
\newblock {\em IEEE Transactions on Medical Imaging}, pages 1--1, 2023.

\bibitem{fengGoldenAngleRadialMRI2022}
Li~Feng.
\newblock Golden-{{Angle Radial MRI}}: {{Basics}}, {{Advances}}, and
  {{Applications}}.
\newblock {\em Journal of Magnetic Resonance Imaging}, 56(1):45--62, 2022.

\bibitem{nattererMathematicsComputerizedTomography2001}
F.~Natterer.
\newblock {\em The Mathematics of Computerized Tomography}.
\newblock Number~32 in Classics in Applied Mathematics. {Society for Industrial
  and Applied Mathematics}, {Philadelphia}, 2001.

\bibitem{el-rewaidyReplicationDataMultiDomain2020}
Hossam {El-Rewaidy}.
\newblock Replication {{Data}} for: {{Multi-Domain Convolutional Neural
  Network}} ({{MD-CNN}}) {{For Radial Reconstruction}} of {{Dynamic Cardiac
  MRI}}, 2020.

\bibitem{el-rewaidyMultidomainConvolutionalNeural2021}
Hossam {El-Rewaidy}, Ahmed~S. Fahmy, Farhad Pashakhanloo, Xiaoying Cai, Selcuk
  Kucukseymen, Ibolya Csecs, Ulf Neisius, Hassan {Haji-Valizadeh}, Bjoern
  Menze, and Reza Nezafat.
\newblock Multi-domain convolutional neural network ({{MD-CNN}}) for radial
  reconstruction of dynamic cardiac {{MRI}}.
\newblock {\em Magnetic Resonance in Medicine}, 85(3):1195--1208, 2021.

\bibitem{blumenthalMrireconBartVersion2022}
Moritz Blumenthal, Christian Holme, Volkert Roeloffs, Sebastian Rosenzweig,
  Philip Schaten, Nick Scholand, Jon Tamir, Xiaoqing Wang, and Martin Uecker.
\newblock Mrirecon/bart: Version 0.8.00.
\newblock Zenodo, September 2022.

\bibitem{ueckerESPIRiTEigenvalueApproach2014}
Martin Uecker, Peng Lai, Mark~J. Murphy, Patrick Virtue, Michael Elad, John~M.
  Pauly, Shreyas~S. Vasanawala, and Michael Lustig.
\newblock {{ESPIRiT}}\textemdash an eigenvalue approach to autocalibrating
  parallel {{MRI}}: {{Where SENSE}} meets {{GRAPPA}}.
\newblock {\em Magnetic Resonance in Medicine}, 71(3):990--1001, 2014.

\bibitem{wangImageQualityAssessment2004}
Zhou Wang, A.C. Bovik, H.R. Sheikh, and E.P. Simoncelli.
\newblock Image quality assessment: From error visibility to structural
  similarity.
\newblock {\em IEEE Transactions on Image Processing}, 13(4):600--612, April
  2004.

\bibitem{waltScikitimageImageProcessing2014}
St{\'e}fan van~der Walt, Johannes~L. Sch{\"o}nberger, Juan {Nunez-Iglesias},
  Fran{\c c}ois Boulogne, Joshua~D. Warner, Neil Yager, Emmanuelle Gouillart,
  and Tony Yu.
\newblock Scikit-image: Image processing in {{Python}}.
\newblock {\em PeerJ}, 2:e453, June 2014.

\bibitem{kingmaAdamMethodStochastic2014}
Diederik~P. Kingma and Jimmy Ba.
\newblock Adam: {{A Method}} for {{Stochastic Optimization}}.
\newblock {\em CoRR}, December 2014.

\bibitem{bradburyJAXComposableTransformations2018}
James Bradbury, Roy Frostig, Peter Hawkins, Matthew James~Johnson, Chris Leary,
  Dougal Maclaurin, George Necula, Adam Paszke, Jake VanderPlas, Skye
  {Wanderman-Milne}, and Qiao Zhang.
\newblock {{JAX}}: Composable transformations of {{Python}}+{{NumPy}} programs,
  2018.

\bibitem{saragadamWIREWaveletImplicit2023}
Vishwanath Saragadam, Daniel LeJeune, Jasper Tan, Guha Balakrishnan, Ashok
  Veeraraghavan, and Richard~G Baraniuk.
\newblock {{WIRE}}: {{Wavelet Implicit Neural Representations}}.
\newblock In {\em Proceedings of the {{IEEE}}/{{CVF Conference}} on {{Computer
  Vision}} and {{Pattern Recognition}}}, pages 18507--18516, 2023.

\bibitem{fengSpatiotemporalImplicitNeural2023}
Jie Feng, Ruimin Feng, Qing Wu, Zhiyong Zhang, Yuyao Zhang, and Hongjiang Wei.
\newblock Spatiotemporal implicit neural representation for unsupervised
  dynamic {{MRI}} reconstruction, January 2023.

\bibitem{kunzImplicitNeuralNetworks2023}
Johannes~F. Kunz, Stefan Ruschke, and Reinhard Heckel.
\newblock Implicit {{Neural Networks}} with {{Fourier-Feature Inputs}} for
  {{Free-breathing Cardiac MRI Reconstruction}}, May 2023.

\end{thebibliography}

\end{document}